\documentclass[reprint,superscriptaddress,amsmath,amssymb,aps,pra,showpacs]{revtex4-2}
\usepackage{graphicx,dcolumn,bm,natbib,xcolor,nicefrac,soul}
\usepackage{hyperref}
\newcommand\HoB{HoB$_4$}
\newcommand\NdB{NdB$_4$}
\newcommand\ErB{ErB$_4$}
\newcommand\TmB{TmB$_4$}
\newcommand\RB{$R$B$_4$}

\newcommand\muB{$\mu_{\mathrm{B}}$}

\newcommand\afm{antiferromagnetic}
\newcommand\TNL{$T_{\mathrm{N3}}=4.9$~K}
\newcommand\TNI{$T_{\mathrm{N2}}=6.8$~K}
\newcommand\TNH{$T_{\mathrm{N1}}=17$~K}

\begin{document}
\title{Field evolution of the multi-$k$ magnetic states in NdB$_4$ revealed by neutron diffraction}
\author{D.~Brunt}
\affiliation{Department of Physics, University of Warwick, Coventry CV4 7AL, United Kingdom}
\affiliation{National Physical Laboratory, Hampton Road, Teddington TW11 0LW, United Kingdom}
\author{N.~Qureshi}
\affiliation{Institut Laue-Langevin, 6 rue Jules Horowitz, BP 156, 28042 Grenoble Cedex 9, France}
\author{E.~Chan}
\affiliation{Institut Laue-Langevin, 6 rue Jules Horowitz, BP 156, 28042 Grenoble Cedex 9, France}
\author{B.~Ouladdiaf}
\affiliation{Institut Laue-Langevin, 6 rue Jules Horowitz, BP 156, 28042 Grenoble Cedex 9, France}
\author{G.~Balakrishnan}
\affiliation{Department of Physics, University of Warwick, Coventry CV4 7AL, United Kingdom}
\author{D.D.~Khalyavin}		\affiliation{ISIS Neutron and Muon Source, STFC Rutherford Appleton Laboratory, Chilton, Didcot, OX11 0QX, United Kingdom}
\author{O.~A.~Petrenko}     \email{petrenkoo@ill.fr}
\affiliation{Department of Physics, University of Warwick, Coventry CV4 7AL, United Kingdom}
\affiliation{Institut Laue-Langevin, 6 rue Jules Horowitz, BP 156, 28042 Grenoble Cedex 9, France}

\date{\today}
\begin{abstract}
We investigate the magnetic states of the Shastry-Sutherland-lattice magnet \NdB\ using single-crystal neutron diffraction.
We focus on magnetic fields applied along the tetragonal $c$~axis, which stabilise the fractional $\nicefrac{1}{5}$~magnetisation plateau.
In the low-temperature phase below 4.9~K, the transition into the plateau state at approximately 12~kOe is marked by a change in the propagation vectors from $(1/5,0,-2/5)$ and its double $(2/5,0,1/5)$ to $(0,0,1/5)$ and $(0,0,2/5)$.
At the transition from the plateau to a nominally `field-polarised' state near 33~kOe, all magnetic reflections associated with non-integer propagation vectors disappear, whereas reflections corresponding to the $\mathbf{k}=0$ component persist throughout the entire field range.
We conclude that that the in-plane components survive even in the high-field phase and that the polarisation is, in fact, incomplete.
The observed field dependence provides direct support for a description of \NdB\ in terms of two coupled magnetic instabilities: all-in-all-out order of the in-plane moments and a spin-density wave of the out-of-plane components.
\end{abstract}
\maketitle

\section{Introduction} \label{sec: Intro}
Rare-earth tetraborides, \RB\ ($R$ is a rare-earth ion), have garnered significant attention due to the diverse magnetic properties observed throughout the entire family of materials.
Their magnetic ground states range from simple \afm\ order found in ErB$_4$~\cite{1979_Pfeiffer_ErB4} and GdB$_4$~\cite{2006_Blanco_GdB4} to more complex states showing the onset of intermediate phases before achieving low-temperature ground state configurations.
These include HoB$_4$, where there is an incommensurate magnetic modulation~\cite{2008_Okuyama_HoB4}, as well as TmB$_4$, which displays a complex set of magnetic phases, some of which are metastable~\cite{2008_Siemensmeyer_TmB4,2009_Michimura_TmB4,2019_Lancon}.

The \RB\ family crystallizes into a tetragonal ($P4/mbm$) crystal structure~\cite{1985_Etourneau_RB4,2001_Salamakha}.
The $R$ ions form a network of squares and triangles in the basal plane, which has been suggested to map to the frustrated Shastry-Sutherland lattice (SSL)~\cite{1981_Shastry,2010_Matas}.
An interesting ingredient of the complex behavior of the \RB\ compounds is their metallic nature, which has been proposed to play a role in the magnetic interactions~\cite{1972_Buschow,1973_Creyghton}.
Fractional magnetisation plateaux are a common feature exhibited by the \RB\ family members and significant theoretical and experimental work has been done to explore their origin~\cite{2009_Moliner, 2013_Huo, 2013_Grechnev, 2012_Huang, 2013_Dublenych,2022_Qureshi}.
The models based on SSL have been successful in describing the field-induced magnetic structures in \ErB~\cite{2013_Wierschem} and \TmB~\cite{2008_Siemensmeyer_TmB4}, however it was later found that the SSL is not well suited to describe the in-field behaviour of \HoB~\cite{2017_Brunt_HoB4}.
In \HoB, the one third magnetisation plateau arises due to stacking of the ferromagnetic layers along the $c$~axis rather than the arrangement of the magnetic moments within the layers, which suggests that significant 3D interactions should be taken into account~\cite{2017_Brunt_HoB4}.

\NdB\ in particular has a complex magnetic phase diagram in applied magnetic field~\cite{2009_Watanuki_NdB4,2018_Brunt_RB4,2023_Ohlendorf}. 
In zero field, it displays successive magnetic phase transitions at \TNH, \TNI, and \TNL~\cite{2009_Watanuki_NdB4} (marginally different transition temperatures, 17.2, 7.0, and 4.8~K were reported in Ref.~\cite{2017_Yamauchi_NdB4} as well as 17.2, 6.8, and 4.8~K in Ref~\cite{2021_Ohlendorf}).
We have previously reported the results of powder neutron diffraction (PND), as well as single-crystal polarised neutron diffraction measurements~\cite{2024_Khalyavin}, which clarified the nature of the magnetic ground state of \NdB.
Here we follow the same conventions as used in Ref.~\cite{2024_Khalyavin} and refer to the magnetic phase formed between $T_{\mathrm{N1}}$ and $T_{\mathrm{N2}}$ as the high temperature (HT) phase, the phase between $T_{\mathrm{N2}}$ and $T_{\mathrm{N3}}$ as the intermediate temperature (IT) phase and the phase appearing below $T_{\mathrm{N3}}$ as the low temperature (LT) phase.

The zero-field LT magnetic structure consists of two separate components.
A representative depiction of the zero-field magnetic structure is given in Fig.~2 of Ref.~\cite{2024_Khalyavin}.
The $\mathbf{k}_1=0$ ``all-in-all-out'' orthogonal spin configuration is confined within the $ab$ plane, with the magnetic moments pointing along the $[110]$ or $[1\overline{1}0]$ directions.
The amplitude of the in-plane Fourier component $m_{xy}^{k_1}=\sqrt{(m_x^{k_1})^2+ (m_y^{k_1})^2}$ is refined to be 1.603(4)~\muB~\cite{2024_Khalyavin}.
The out-of-plane component points along the $c$~axis and forms an anharmonic spin density wave characterised by the propagation vectors $\mathbf{k}_1=0$,  $\mathbf{k}_2= (1/5,0,-2/5)$, and $\mathbf{k}_3=2\mathbf{k}_2=(2/5,0,-4/5).$
The corresponding Fourier components are refined to be $m^{k1}_z=0.925(3)$~\muB, $m^{k2}_z=1.719(7)$~\muB, and $m^{k3}_z=0.656(3)$~\muB~\cite{2024_Khalyavin}.

The commensurate $\mathbf{k}_1=0$ structure is also present in the IT and HT phases~\cite{2018_Metoki_NdB4,2024_Khalyavin}.
The zero-field IT phase shows the presence of incommensurate reflections characterised by the propagation vector $\mathbf{k}_\mathrm{IT}=(0.142,0.142,2/5)$.
Remarkably, the in-plane and out-of-plane components follow different temperature dependence within the HT phase~\cite{2024_Khalyavin}.
 
Magnetisation measurements in the LT phase reveal the presence of a magnetisation plateau for $H\parallel c$~\cite{2018_Brunt_RB4}.
Analogous to the fractional plateaux observed in other \RB\ compounds, this plateau is well defined and relatively wide, extending approximately from 12 to 33~kOe at $T=2.0$~K.
However, both transition fields are temperature dependent and exhibit significant hysteresis, particularly at the lower-field transition~\cite{2018_Brunt_RB4,Brunt_thesis,2023_Ohlendorf}.
The magnetisation on the plateau is one-fifth of the saturation magnetisation, with $M_{\mathrm{50 kOe}}$ taken as a nominal saturation value (although the magnetisation at this field is substantially lower than expected for the Nd$^{3+}$ ions and it continues to increase steadily in fields above 50~kOe~\cite{2018_Brunt_RB4}).

In this paper, we follow the development of the complex magnetic order in \NdB\ under an applied field using single-crystal neutron diffraction measurements. 
Upon entering the fractional $\nicefrac{1}{5}$~magnetisation plateau phase, the LT zero-field reflections associated with the propagation vectors $\mathbf{k}_2$ and $\mathbf{k}_3$ are replaced by the reflections characterised by $\mathbf{k}_{\mathrm 1/5}=(0,0,1/5)$ and 2$\mathbf{k}_{\mathrm 1/5}$.
At the transition to the high-field phase at about 33~kOe, all non-integer magnetic reflections disappear, leaving only magnetic reflections with $\mathbf{k}=0$.
We provide a detailed description of the field evolution of the out-of-plane Fourier components.
Furthermore, analysis of the diffraction data for the integer magnetic peaks reveals that the in-plane magnetisation does not vanish at 33~kOe but persists to higher applied fields.
This observation demonstrates that the high-field magnetic phase above 33~kOe is not fully field-polarised.
This conclusion is consistent with previous high-field magnetisation measurements~\cite{2018_Brunt_RB4}, which demonstrated a continuous steady increase in the magnetisation for $H \parallel c$, from 2.0~$\mu_{\rm B}$ per Nd ion at 100~kOe to 2.3~$\mu_{\rm B}$ at 500~kOe.

\section{Experimental Details} \label{Sec_Exp_Details}
\NdB\ crystals were grown using the floating zone technique.
Isotopically enriched boron, $^{11}$B (99$\%$) was used to reduce neutron absorption.
All details on crystal growth and sample characterisation can be found in Ref.~\cite{2019_Brunt_Crystal}.

Neutron diffraction measurements were performed at the Institut Laue-Langevin (ILL), Grenoble, France, and were conducted in two stages~\cite{D10_2014,D10_2016,D10_2026}.

In the first stage, single-crystal neutron diffraction experiments were carried out on the D10 diffractometer, both in zero field and in a vertical-field cryomagnet.
A two dimensional 80$\times$80~mm$^2$ area detector was used for all measurements~\cite{1988_Wilkinson}.
A 4-circle setup with a closed-cycle cryostat with a Joule-Thomson stage was used for the zero field measurements, while an asymmetric cryomagnet with vertical opening of $+20^{\circ}$ and $-5^{\circ}$ supplying a magnetic field of up to 60~kOe was used for measurements in the applied field.
The sample (0.33~g) was fixed to an aluminium holder and the magnetic field was applied along the vertical $c$~axis defining the horizontal $(hk0)$ scattering plane.
The accuracy of the alignment of the $c$~axis of the crystal along the magnetic field was better than $1^{\circ}$.
Incident wavelengths of $\lambda = 2.36$~\AA\ and $\lambda = 1.26$~\AA\ were obtained using pyrolytic graphite (PG) and Cu(200) monochromators, respectively.
A PG filter was used to reduce contamination of $\lambda/2$ in the case of $\lambda = 2.36$~\AA.
The shorter wavelength allowed measurement of incommensurate reflections outside of the horizontal scattering plane.

\begin{figure}[tb]
\includegraphics[width=0.995\columnwidth]{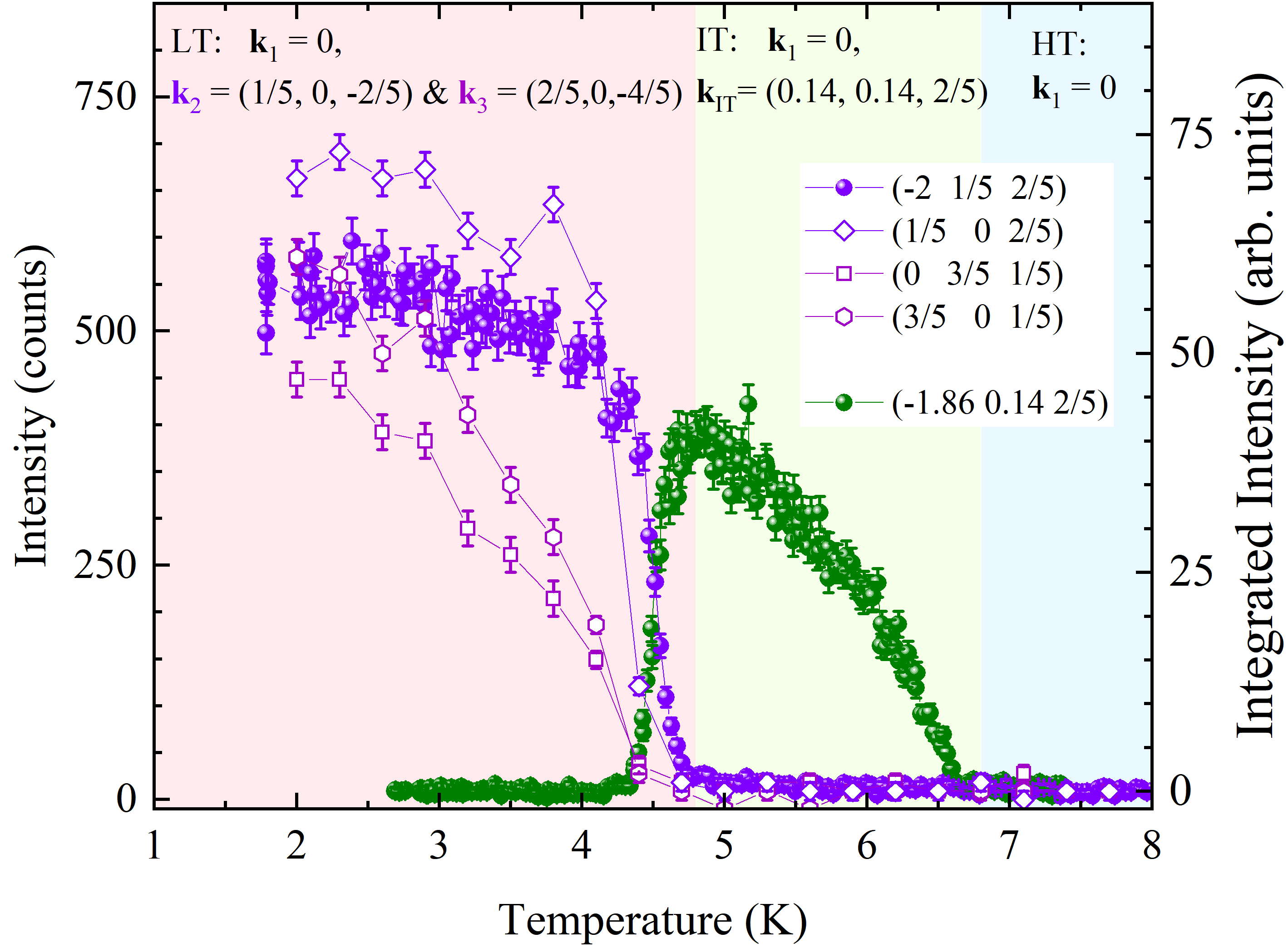}
\caption{Temperature dependence of the intensity of selected incommensurate (IT) and fractional (LT) reflections in \NdB\ measured in zero magnetic field.
Solid symbols correspond to continuous measurements of the peak intensity acquired while slowly varying the temperature (left-hand intensity scale).
Open symbols represent integrated intensities obtained from scans through the reflections at fixed temperatures (right-hand intensity scale).}
\label{fig:incomm_scans}
\end{figure}

For the second part of the study, neutron diffraction measurements in a horizontal magnetic field were performed several years later.
By that time, the diffractometer had been upgraded to a new configuration, designated D10+.
The rebuild and re-installation of a new neutron guide and a new vertically focusing monochromator resulted in a substantial increase in neutron flux.
A different single crystal from the same growth batch (mass = 0.45~g) was used for this part of the experiment, preventing a direct comparison of Bragg peak intensities between the two experimental stages.
An incident wavelength of $\lambda =2.38$~\AA, selected using PG monochromator was employed together with a PG filter suppressing higher-order contaminations.
The magnetic field was provided by a horizontal-field cryomagnet with a maximum field of 38~kOe.
As in the first experiment, the magnetic field was applied parallel to the crystallographic $c$ axis; however, in this configuration the horizontal scattering plane was $(h0l)$.
The accuracy of the field alignment was more difficult to estimate, but the consistency of the results indicates that the misalignment was less than $2^{\circ}$.

For both parts of the experiment, field- and temperature-dependent intensities were measured by summing the neutron counts within a small region of the detector centred on the reflection of interest.
Transverse and longitudinal scans through selected reflections were also recorded at fixed field and temperature.

\section{Neutron Diffraction Results} \label{Sec_Results}
\subsection{Zero field} \label{subsec_0}

Figure~\ref{fig:incomm_scans} shows the temperature dependence of the intensity of a selection of fractional reflections in the LT phase and an incommensurate reflection of the IT phase and in zero field.
The background colours in Fig.~\ref{fig:incomm_scans} distinguish the LT, IT, and HT magnetic phases.
The incommensurate reflections observed in the IT phase and the fractional reflections observed in the LT phase are consistent with previous PND results~\cite{2024_Khalyavin}.
The corresponding propagation vectors are $\mathbf{k}_{\mathrm{IT}}=(0.14,0.14,2/5)$ and $\mathbf{k}_2=(1/5,0,-2/5)$, respectively, while the LT phase also exhibits the second harmonic at $(2/5,0,1/5)$.
In the HT phase, only $\mathbf{k}=0$ reflections are observed.

\subsection{Vertical magnetic field}
\label{subsec_vert}
\begin{figure}[tb]
\includegraphics[width=0.995\columnwidth]{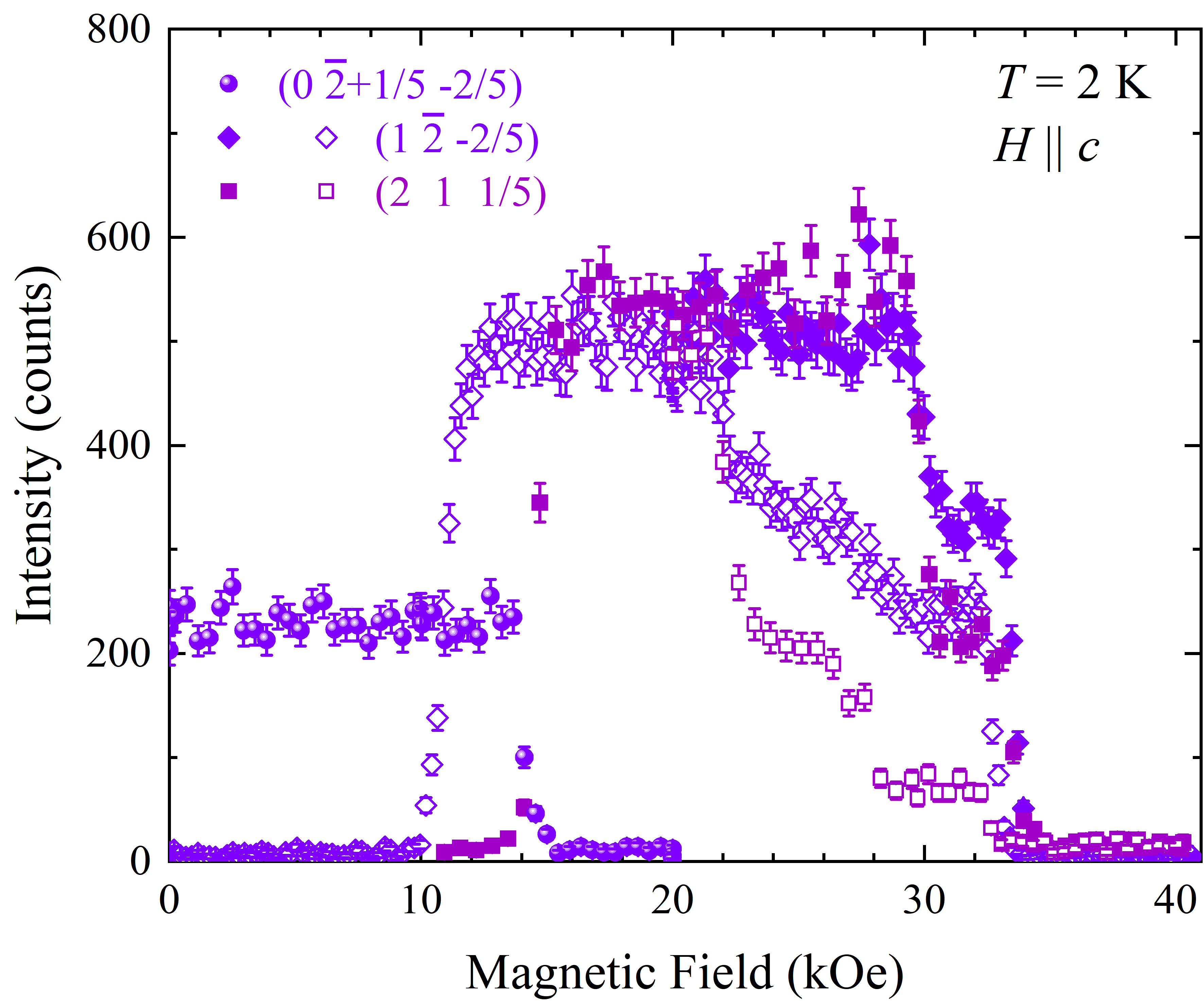}
\caption{Magnetic-field dependence of the intensities of three fractional reflections in the LT phase.
Circles denote the $\mathbf{k}_2$ reflections, while other symbols represent the $\mathbf{k}_{\mathrm{1/5}}$ reflections.
Filled and open symbols correspond to measurements performed with increasing and decreasing magnetic field, respectively.}
\label{fig:Frac_in_field}
\end{figure}

Figure~\ref{fig:Frac_in_field} illustrates the field evolution of the fractional $(hkl)$ peaks in the LT phase.
By scanning through various fractional $(hkl)$ peaks we found that they remain at the same positions in reciprocal space and that only their intensity changes as a function of an applied field.
In the field below the $\nicefrac{1}{5}$~plateau, the intensity of the fractional peaks with propagation vectors $\mathbf{k}_2$ and $\mathbf{k}_3$ (such as the peak $(0~~\overline{2}\!+\!1/5~-\!2/5)$ for example) remains practically constant.
On entering the plateau, these peaks disappear and they are replaced by peaks such as $(1~\overline{2}~-\!2/5)$ and $(2~1~1/5)$.
The new field-induced fractional peaks are described by the propagation vector $\mathbf{k}_{\mathrm 1/5}=(0,0,1/5)$ and its double $(0,0,2/5)$.
All the fractional $(hkl)$ peaks disappear completely on entering the high-field state above 33~kOe.

The measurements reveal significant hysteresis around the $\nicefrac{1}{5}$-magnetisation plateau as well as considerably different intensity of the same peaks depending on whether the field was increasing or decreasing.
Additional smaller step-like features in intensity are clearly visible for the $\mathbf{k}_{\mathrm 1/5}$ peaks within the plateau region.
Previous high-resolution dilatometry studies, combined with the magnetisation measurements of \NdB\ suggested the presence of a narrow $\nicefrac{1}{4}$-magnetisation plateau in addition to a wider $\nicefrac{1}{5}$~plateau.
However, we have not detected the presence of the additional $(0,0,1/4)$-type peaks in our measurements, nor have we observed any appreciable changes in the positions of the fractional $(hkl)$ peaks.
Therefore the neutron diffraction data seem to be more consistent with the notion of metastable states and incomplete transition between the $\nicefrac{1}{5}$~plateau and the field-polarised phases (not uncommon in frustrated magnets) resulting in an {\it accidental} $\nicefrac{1}{4}$-magnetisation value rather than with a stabilisation of an additional fractional plateau of a distinct periodicity. 
We will return to this point in the next subsection when presenting the diffraction results for a horizontally applied field.

\begin{figure}[tb]
\includegraphics[width=0.995\columnwidth]{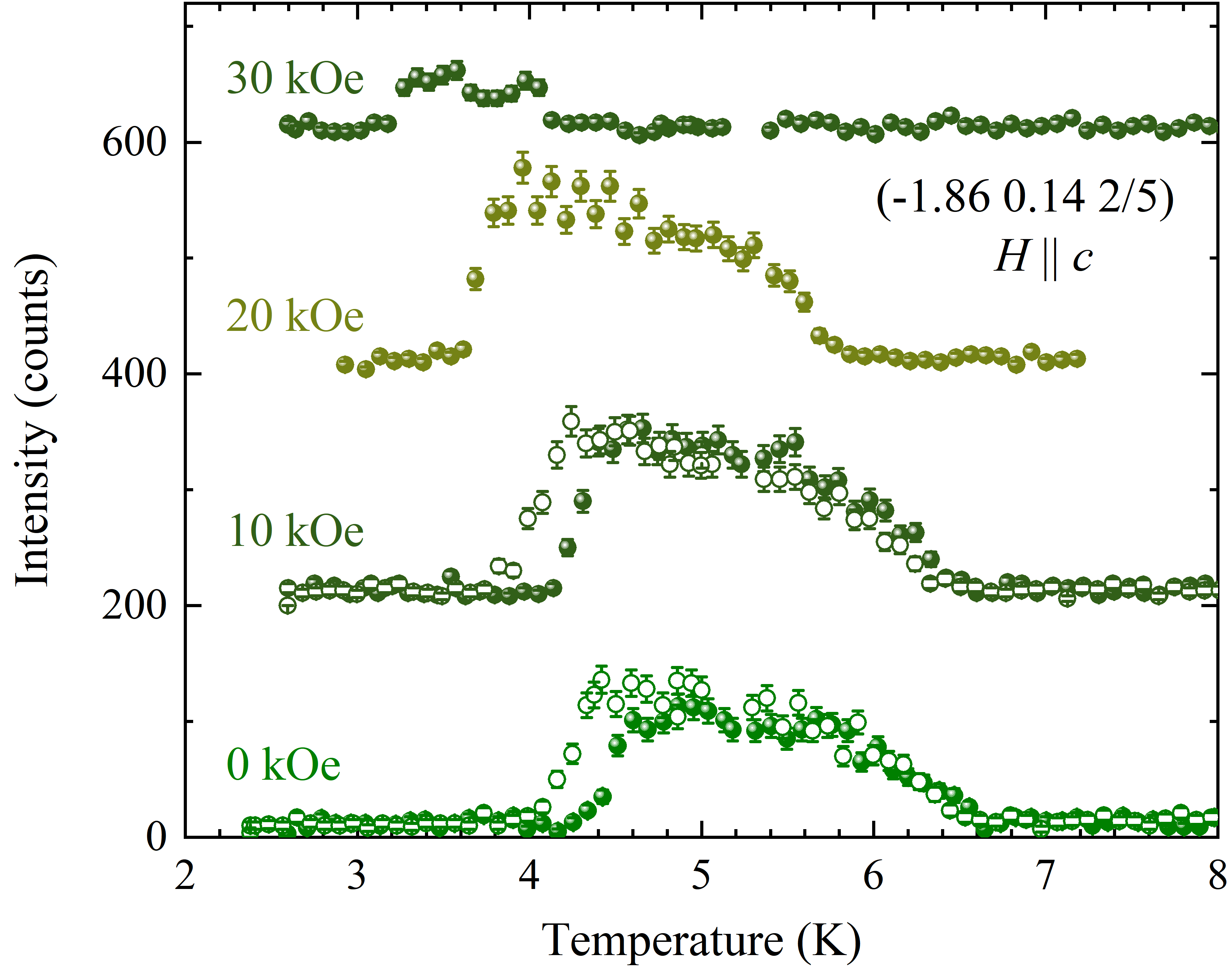}
\caption{Temperature dependence of the intensity of the IT-phase reflection $(-1.86~0.14~2/5)$ measured in different applied magnetic fields.
Closed and open symbols correspond to measurements acquired on warming and cooling, respectively.
For clarity, successive curves are offset by 200 counts.}
\label{fig:Incom_in_field}
\end{figure}

Figure~\ref{fig:Incom_in_field} shows the temperature dependence of the intensity of the incommensurate $(-1.86~0.14~2/5)$ reflection measured in different applied fields.
This reflection is associated with the propagation vector $\mathbf{k}_\mathrm{IT}$ and exists only within the IT phase.
A pronounced hysteresis around $T_{\mathrm{N3}} = 4.8$~K is consistent with specific-heat measurements~\cite{2009_Watanuki_NdB4,Brunt_thesis}, which likewise indicate that this transition is first order.
Upon application of a magnetic field, the temperature range over which the reflection is observed shifts to lower temperatures.
At 10~kOe, the reflection is present between 4.1 and 6.2~K, while at 20~kOe the range is further reduced to 3.6-5.8~K.
At 30~kOe, the intensity of the incommensurate reflection decreases dramatically, and it survives only between 3.1 and 4.1~K.
The behaviour of these and other incommensurate peaks (not shown) is fully consistent with the absence of a fractional magnetisation plateau in the IT phase.
Furthermore, the suppression of the $\mathbf{k}_\mathrm {IT}$ order in an applied magnetic field is also consistent with the $H-T$ phase diagram inferred from bulk property measurements~\cite{2018_Brunt_RB4}. 

\begin{figure}[tb]
\includegraphics[width=0.995\columnwidth]{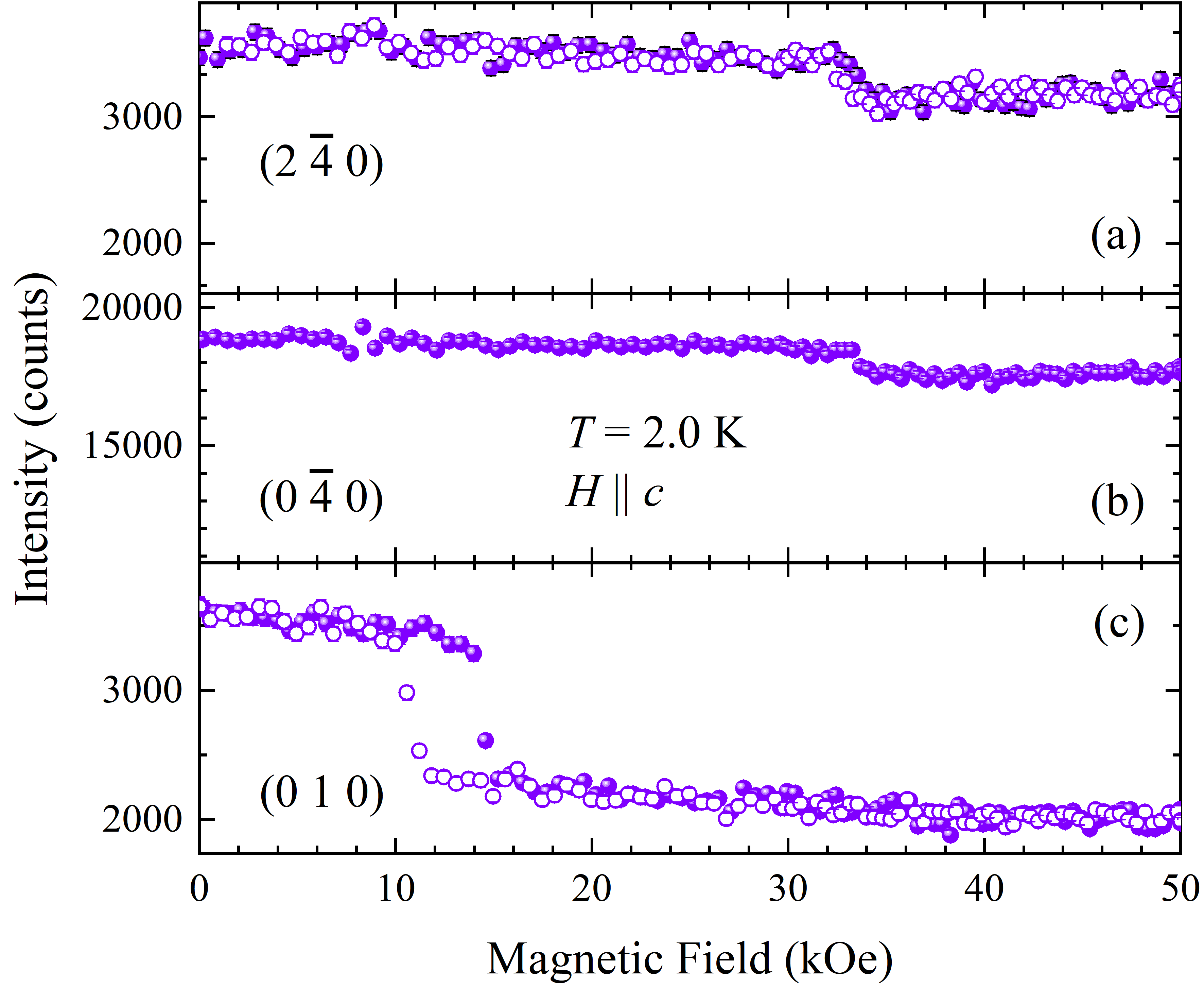}
\caption{Magnetic-field dependence of the intensity of the (a) $(2\overline{4}0)$, (b) $(0\overline{4}0)$, and (c) $(010)$ reflections measured at 2~K.
Closed and open symbols correspond to measurements performed with increasing and decreasing magnetic field, respectively.
The zero level of each $y$ axis was set to the intensity of the corresponding nuclear Bragg reflection in order to emphasise the magnetic contribution to the scattering.}
\label{fig:hk0_field}
\end{figure}

Figure~\ref{fig:hk0_field} shows the field dependence of the intensity of a selection of Bragg reflections with integer indices.
For all plots in Fig.~\ref{fig:hk0_field}, the lower limit of the $y$ axis is set to the intensity of the nuclear Bragg peak contribution measured above \TNH\ in the paramagnetic regime.
Consequently, the displayed intensity variation reflects predominantly the magnetic contribution to the peak intensity.
At 2.0~K, the $(2\overline{4}0)$ and $(0\overline{4}0)$ reflections show a noticeable decrease in intensity only upon entering the high-field phase at approximately 33~kOe.
In contrast, the transition to the $\nicefrac{1}{5}$-magnetisation plateau at 12~kOe has no visible effect on the intensity of either reflection.
In the highest applied field of 50~kOe, the reduction in the magnetic contribution to the intensity remains relatively small, amounting to approximately 10\% and 5\% for the $(2\overline{4}0)$ and $(0\overline{4}0)$ reflections, respectively. 
The (010) reflection, on the other hand, demonstrates an abrupt decrease in intensity at approximately 12~kOe upon entering the $\nicefrac{1}{5}$-magnetisation plateau regime.
Again, this transition is accompanied by a large hysteresis pointing to its first-order nature.
An increase in the applied field from 12 to 50~kOe results in a further, nearly linear decrease in the intensity of the (010) peak.
In the high-field phase, the magnetic component if this peak becomes very weak, with the total intensity only slightly exceeding the nuclear contribution.

By measuring the field dependence of the symmetry equivalent reflections, we have confirmed that the reduction in the intensity of the $(2\overline{4}0)$ peak is not associated with a simple repopulation of magnetic domains, as all ($\pm$2~$\pm$4~0) and ($\pm$4~$\pm$2~0) peaks showed similar behaviour.
Further checks were made for the ($\pm$4~0~0), (0~$\pm$4~0), ($\pm$1~0~0), and (0~$\pm$1~0) peaks and led to the same conclusion for the peaks $(0\overline{4}0)$ and (010).

\begin{figure}[tb]
\includegraphics[width=0.95\columnwidth]{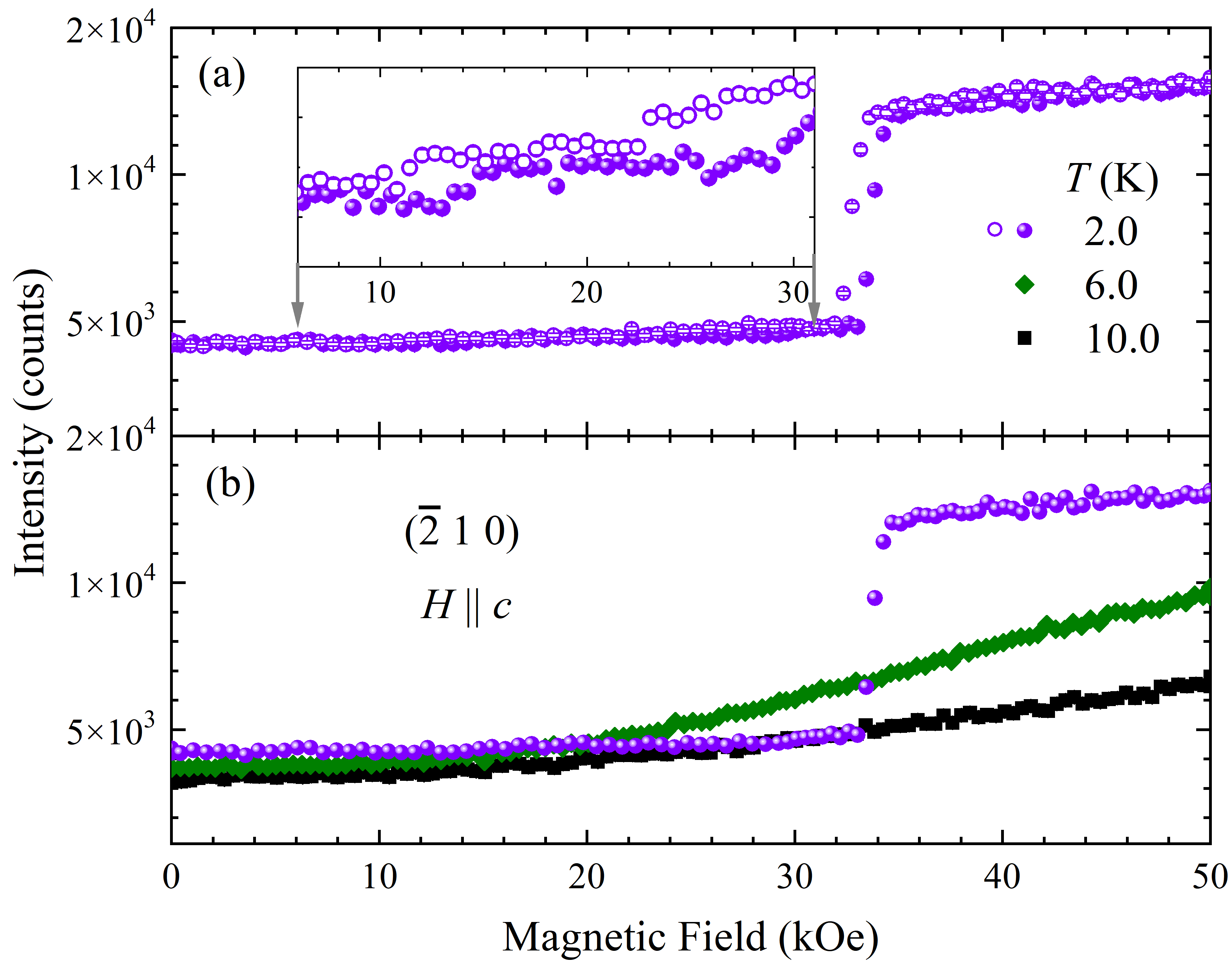}
\caption{(a) Magnetic-field dependence of the intensity of the $(\overline{2}10)$ reflection measured at 2~K.
The inset highlights additional sharp changes in the peak intensity observed within the $\nicefrac{1}{5}$-magnetisation plateau when the field scan was performed with longer counting times.
Closed and open symbols correspond to measurements acquired with increasing and decreasing magnetic field, respectively.
(b) Comparison of the field dependence of the $(\overline{2}10)$ reflection at 2, 6, and 10~K, corresponding to the LT, IT, and HT phases, respectively.}
\label{fig:210_vs_field}
\end{figure}

Figure~\ref{fig:210_vs_field} demonstrates the field dependence of the $(\overline{2}10)$ peak in the LT, IT and HT phases.
The $(\pm 2 \pm 1~0)$ and $(\pm 1 \pm 2~0)$ peaks have a relatively small nuclear component (around 27\% of the total intensity in zero field), therefore, the field dependence of their magnetic component is the easiest to follow.
In the HT phase, the intensity of the $(\overline{2}10)$ reflection exhibits only a marginal increase with increasing magnetic field.
In the IT phase, the increase is considerably more pronounced, although it remains smooth and continuous throughout the field range studied.
Only in the LT phase does the transition at approximately 33~kOe become clearly defined.
At 50~kOe, the magnetic contribution to the intensity is approximately four times larger than in zero field.
As discussed above, the increase in the intensity of the $(\overline{2}10)$ reflection cannot be attributed simply to a redistribution of magnetic domain populations.

Apart from the sharp transition at 33~kOe accompanied by a significant hysteresis, a more detailed inspection (see the inset in Fig.~\ref{fig:210_vs_field}(a)) reveals appreciable variation of the intensity of the $(\overline{2}10)$ peak in the magnetisation plateau region, also accompanied by hysteresis.
The phenomenon of further splitting of fractional magnetisation plateau into a collection of shorter plateaux is a common feature in frustrated magnetism at very low temperatures, it has been seen particularly pronounced in $\rm Ca_3Co_2O_6$, for example ~\cite{2004_Hardy}.

Although the D10 diffractometer was equipped with a lifting arm that allowed the detector to be positioned outside the horizontal scattering plane, the limited vertical opening of the cryomagnet severely restricted the range of accessible reflections.
To access integer $(h0l)$ reflections, which provide crucial information on the field dependence of the in-plane magnetic component, the experimental setup was therefore reconfigured to use a horizontal-field cryomagnet.

\subsection{Horizontal magnetic field}
\label{subsec_hor}
\begin{figure}[tb]
\includegraphics[width=0.995\columnwidth]{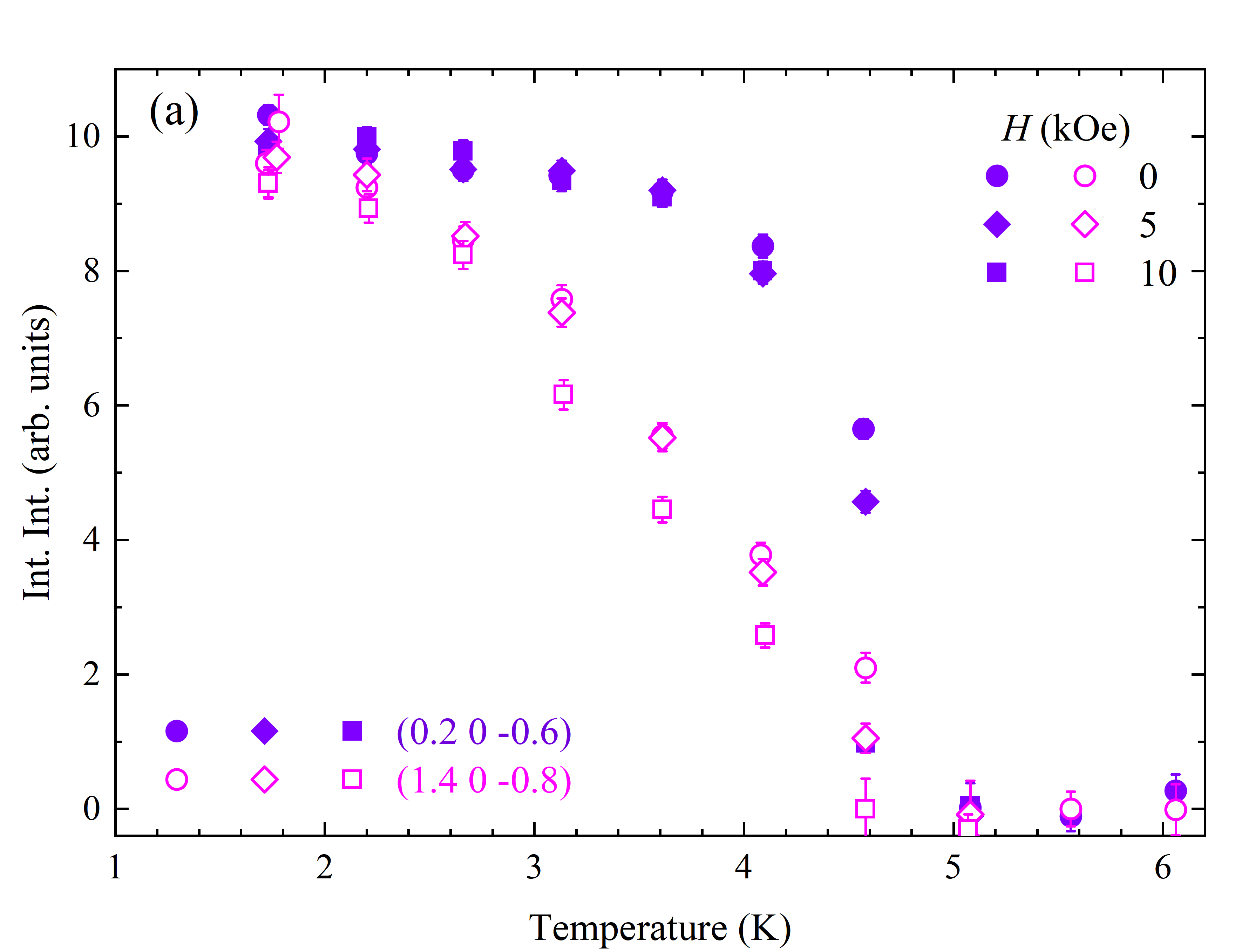}
\includegraphics[width=0.995\columnwidth]{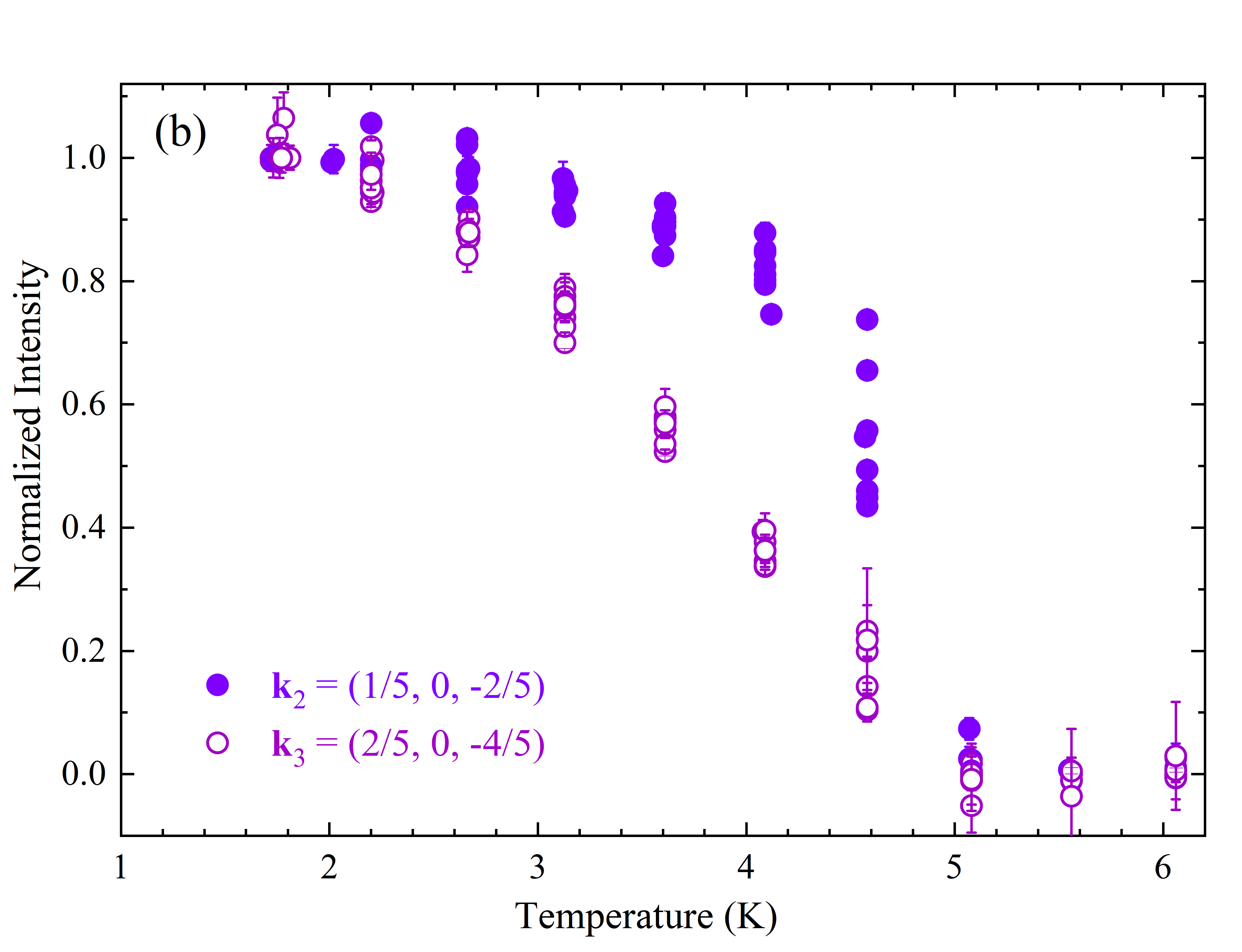}
\caption{(a) Temperature dependence of the intensity of the (0.2~0~0.6) and (1.4~0~-0.8) reflections, representative of the $\mathbf{k}_2$ and $\mathbf{k}_3$ propagation vectors, respectively, in applied fields of 0, 5, and 10~kOe.
(b) Normalised intensity of several $\mathbf{k}_2$ (solid symbols) and $\mathbf{k}_3$ (open symbols) reflections, illustrating their different temperature dependences.
The panel shows only the 0 and 5~kOe datasets.}
\label{fig:k2k3_vs_T}
\end{figure}

The temperature dependence of the intensity of the (0.2~0~0.6) and (1.4~0~-0.8) reflections was measured in zero field as well as in 5 and 10~kOe, see Fig.~\ref{fig:k2k3_vs_T}(a).
These peaks are characterised by the propagation vectors $\mathbf{k}_2$ and $\mathbf{k}_3$ respectively.
In zero field, both the $\mathbf{k}_2$ and $\mathbf{k}_3$ peaks disappear on warming above  $T_{\mathrm {N3}}$ and the field dependence appears rather weak.
The applied magnetic fields of 5 and 10~kOe do not significantly moderate the behaviour of the fractional peaks at low $T$, but at higher $T$, on approaching a transition, the intensity in 10~kOe field seems to be lower compared to 0 and 5~kOe. 
Thus  the $T_{\mathrm {N3}}$ shows only a very marginal field dependence (a small reduction).

In Fig.~\ref{fig:k2k3_vs_T}(b) we combine normalised intensity of four different peaks from each families of the $\mathbf{k}_2$ and $\mathbf{k}_3$ propagation vectors to emphasise their different temperature dependence.
For clarity, only the 0 and 5~kOe data are shown in panel (b).
Apart from the two reflections shown in panel (a), we have also measured (0.2~0~0.4), (0.2~0~-0.6), (-0.2~0~-0.6), (0.6~0~0.2), (0.6~0~-0.2) and (0.6~0~-0.8).
The $\mathbf{k}_3$ reflections show a stronger temperature dependence than the fundamental $\mathbf{k}_2$ peaks.
This observation, fully consistent with the data shown in Fig.~\ref{fig:incomm_scans}, indicates a progressive reduction of the second-harmonic content of the magnetic structure on warming.
The magnetic modulation seems to evolve towards a perfectly sinusoidal form at $T_{\mathrm {N3}}$ -- the low-$T$ squared-up character of the magnetic structure is gradually loss on warming.

\begin{figure}[tb]
\includegraphics[width=0.995\columnwidth]{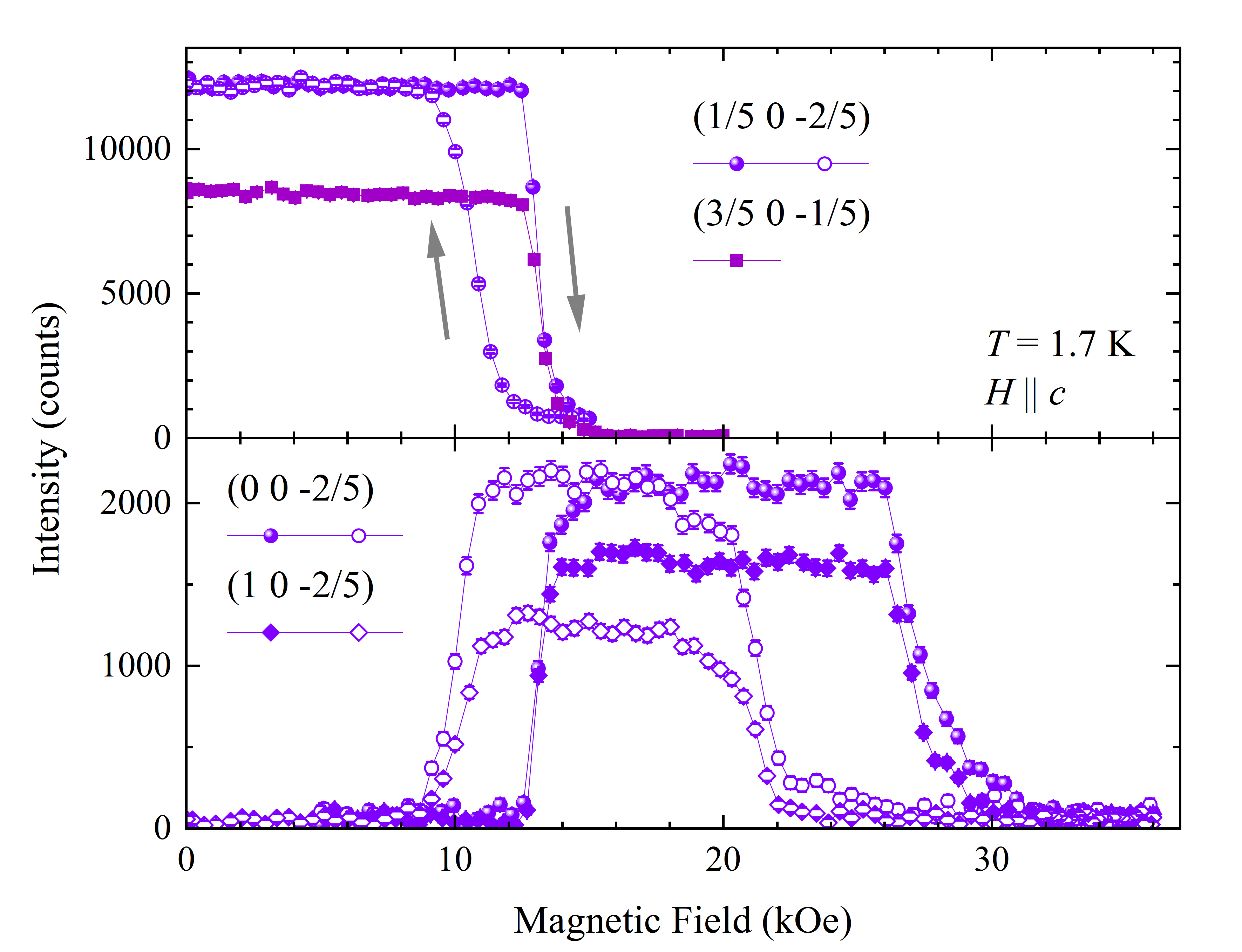}
\caption{Magnetic-field dependence of the intensity of fractional reflections in the LT phase measured in horizontal field.
Closed and open symbols correspond to measurements performed with increasing and decreasing magnetic field, respectively.}
\label{fig:Frac_in_horizontal_field}
\end{figure}

Figure~\ref{fig:Frac_in_horizontal_field} the magnetic-field dependence of the intensities of the fractional reflections in the LT phase for a field applied within the horizontal scattering plane.
In agreement with the measurements performed in a vertical field, the $\mathbf{k}_2$ and $\mathbf{k}_3$ type reflections are observed at low fields and are replaced by $\mathbf{k}_{\mathrm 1/5}$ reflections upon entering the $\nicefrac{1}{5}$~magnetisation plateau.
The $\mathbf{k}_{\mathrm 1/5}$ reflections, in turn, disappear on entering the high-field phase.
A pronounced hysteresis is observed for all fractional reflections at both the lower- and higher-field boundaries of the plateau.
The slightly different high-field transition fields observed for vertical and horizontal field orientations may arise from the substantially different demagnetisation factors of the samples used in the two measurements.

\begin{figure}[tb]
\includegraphics[width=0.995\columnwidth]{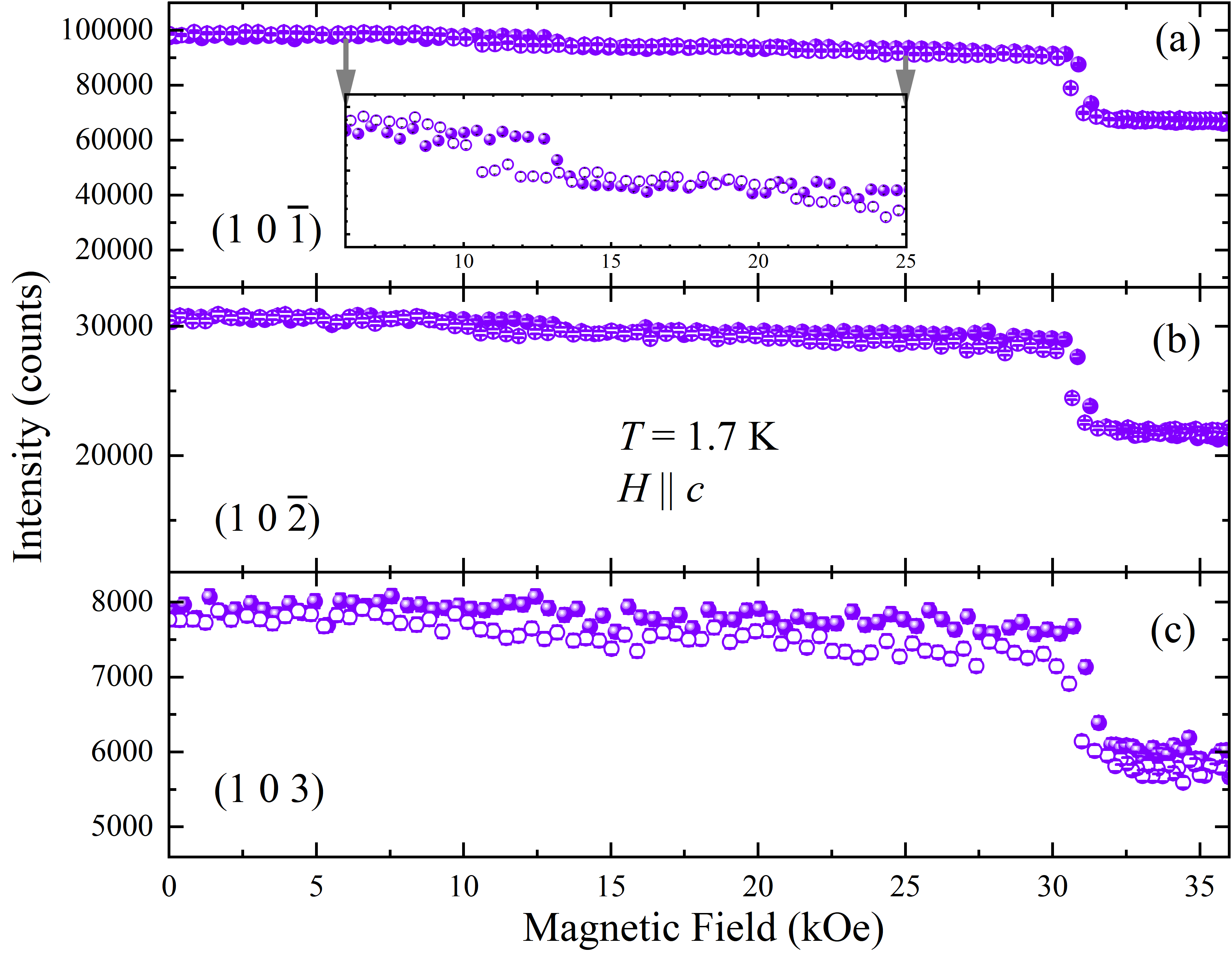}
\caption{Magnetic-field dependence of the intensity of integer $(h0l)$ reflections measured at 1.7~K in a horizontal applied field.
Closed and open symbols correspond to measurements performed with increasing and decreasing magnetic field, respectively.
The zero level of each $y$ axis was set to the intensity of the corresponding nuclear Bragg reflection in order to emphasise the magnetic contribution to the scattering.
The inset in panel (a) illustrates variation of the intensity of the $(1 0 \bar{1})$ peak on the $\nicefrac{1}{5}$-magnetisation plateau.}
\label{fig:h0l_field}
\end{figure}

The most important results of the measurements performed in a horizontal field are presented in Fig.~\ref{fig:h0l_field}.
In a manner similar to what is done in Fig.~\ref{fig:hk0_field}, the zero levels of $y$~axis are set to the intensity of the corresponding nuclear Bragg reflection.
The choice of particular reflections to follow was largely influenced by their accessibility in a hotizontal field magnet. 
The magnetic intensity of the integer $(10l)$-type reflections decreases slightly on entering the $\nicefrac{1}{5}$-magnetisation plateau and is further strongly suppressed upon transition into the field-polarised phase.
Nevertheless, the reflections remain clearly observable above the upper critical field, demonstrating that a substantial component of the in-plane magnetic structure survives in the high-field state.

\section{Discussion and Summary} \label{Sec_Conclusions}
\begin{figure*}[tb]
\includegraphics[width=1.99\columnwidth]{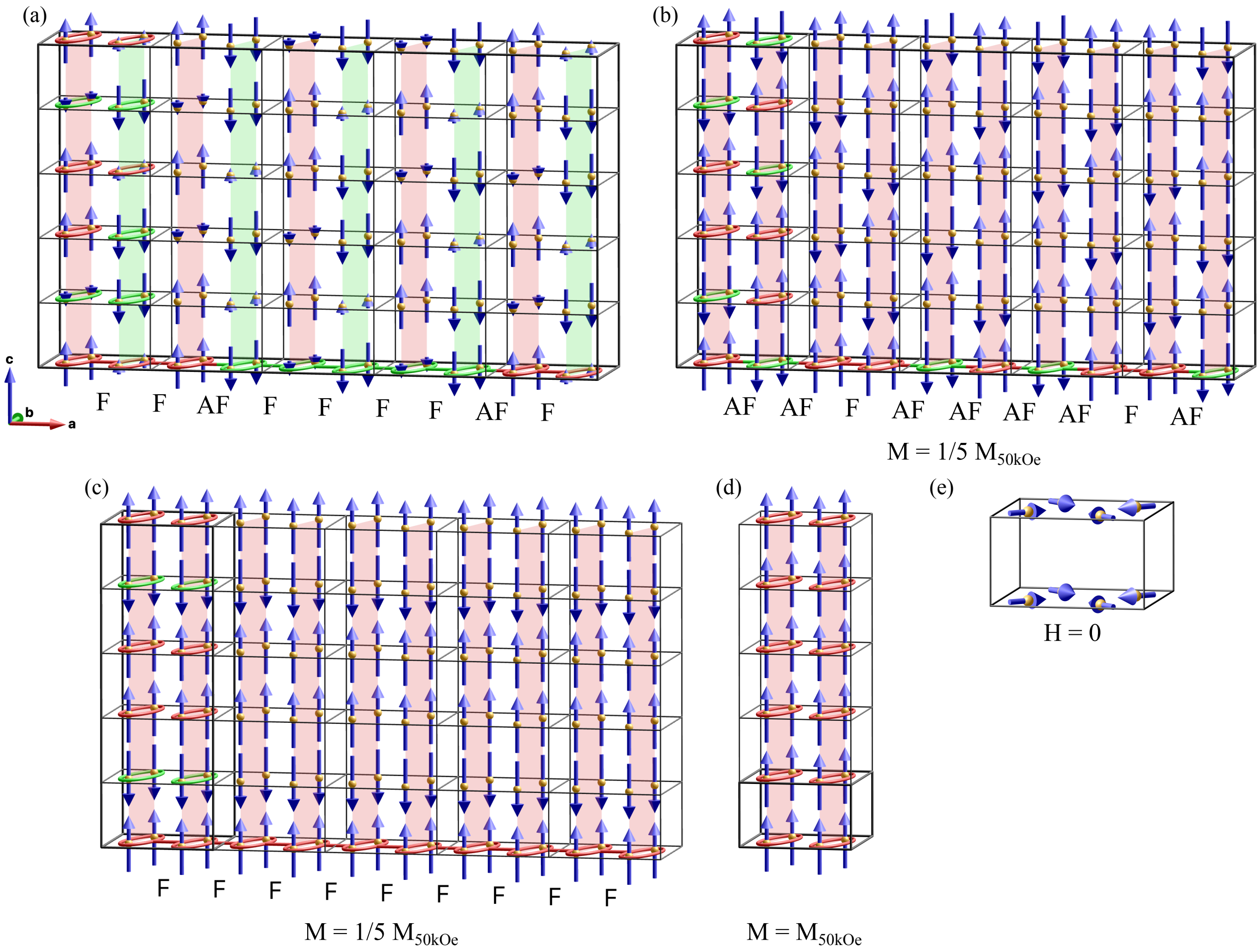}
\caption{
(a) The out-of-plane components of the Nd$^{3+}$ magnetic moments of the zero-field LT structure of \NdB.
The structure is three-up-two-down order with fivefold periodicity along $a$ and $c$.
The moments are \afm ally stacked along $a$~axis ferromagnetic $bc$~planes are described by the $m^{k1}_z$ component, the internal field is staggered (depicted in red and green) and the net magnetisation is zero.
The length of the magnetic moments reflects the results of powder neutron refinements~\cite{2024_Khalyavin}.
(b) Possible magnetic configuration with $\nicefrac{1}{5}$ of the total moment.
It represents an idealised case with equal-moments for all Nd sites.
(b) is identical to (a) but with the directions of the moments of the green $bc$~planes reversed.
Propagation vectors for (b) are the same as in zero field, $\mathbf{k}_2= (1/5,0,-2/5)$, and $\mathbf{k}_3=2\mathbf{k}_2=(2/5,0,-4/5)$.
This configuration is not found experimentally.
(c) The magnetic structure adopted by \NdB\ in the region of $\nicefrac{1}{5}$-magnetisation plateau.
The equal-moment structure is still \textsc{uuudd} along the $c$~axis but stacking along $a$ is ferromagnetic, propagation vector is $\mathbf{k}_{\mathrm 1/5}=(0,0,1/5)$.
(d) The out-of-plane components of the Nd$^{3+}$ magnetic moments in a field-polarised phase.
(e) The in-plane components in zero field.
The figure was created using Mag2Pol~\cite{mag2pol}.}
\label{fig:MagStructure}
\end{figure*}

We have made a detailed investigation of the magnetic structure of \NdB\ in an applied magnetic field using neutron diffraction techniques.
Unfortunately, due to large extinction and significant multiple scattering effects occurring in the single crystal samples we are unable to perform a full in-field refinement to determine the exact magnetic structure.
However, we are able to explain the observed field-induced changes in the intensity of magnetic Bragg peaks within a recently proposed concept of coupled, coexisting magnetic instabilities, the all-in-all-out order of the in-plane components of the Nd moments and the anharmonic spin density wave of the out-of-plane components.
The latter represents 'three-up-two-down' (\textsc{3u2d}) and 'three-down-two-up' (\textsc{3d2u}) spin arrangements within the $bc$~planes, alternating along the $a$~axis, which is the observed LT zero-field magnetic structure (see Fig.~\ref{fig:MagStructure}(a)).
Swapping the \textsc{3d2u} planes to \textsc{3u2d} ones, should inevitably result in the magnetic structure with an uncompensated net magnetic moment, naturally explaining the $\nicefrac{1}{5}$~plateau observed in the magnetisation measurements.
The alternating \textsc{3u2d}/\textsc{3d2u} sequence has been suggested to be imposed by the internal staggered field, originating from a coupling between the in plane all-in-all-out spin component and a non-magnetic ``hidden'' order parameter~\cite{2024_Khalyavin}.

The Fourier decomposition of the anharmonic spin density wave consists of three components, $m_z^{k1}$, $m_z^{k2}$, and $m_z^{k3}$, with $\mathbf{k}_1=0$,  $\mathbf{k}_2= (1/5,0,-2/5)$, and $\mathbf{k}_3=2\mathbf{k}_2=(2/5,0,-4/5)$ propagation vectors, respectively.
The $m_z^{k1}$ component follows the internal staggered field and represents the alternation of the net magnetisation within the \textsc{3u2d}/\textsc{3d2u} $bc$~planes.
The magnetic field-induced transition to the uniform \textsc{3u2d} configuration, should therefore result in the replacement of the \afm\ $m_z^{k1}$ Fourier component with a ferromagnetic one.
The reflection sensitive to the former is (010).
This is a structurally forbidden reflection with no intensity contribution from the ferromagnetic spin alignment.
As seen from Fig.~\ref{fig:hk0_field}(c), the intensity of this reflection drastically decreases in the $\nicefrac{1}{5}$~plateau phase.
On entering the plateau, the (010) reflection does not completely vanish, but this is likely due to multiple scattering, where a finite intensity is also observed at high temperatures in the paramagnetic phase.
This reflection has not been observed in previous powder diffraction experiment~\cite{2024_Khalyavin} in compliance with the requirements of the $P4/mbm$ space group.

The simplest scenario for the field-induced transition from the \textsc{3u2d}/\textsc{3d2u} state to a uniform \textsc{3u2d} spin configuration does not require any change in the $\mathbf{k}_2$ and $\mathbf{k}_3$ propagation vectors.
The corresponding model, shown in Fig.~\ref{fig:MagStructure}(b), can be obtained by reversing the spins in half of the $bc$~planes, following the staggered component of the internal field that opposes the externally applied field (green planes in Fig.~\ref{fig:MagStructure}(a)).
In Fig.~\ref{fig:MagStructure}(a)-(c), ferromagnetic and \afm\ alignments of the Nd$^{3+}$ moments along the $a$~axis are denoted by F and AF, respectively.
The model shown in Fig.~\ref{fig:MagStructure}(b) implies that $\nicefrac{4}{5}$ of the nearest-neighbour couplings $J^z_1$ (using the notation of Ref.~\cite{2024_Khalyavin}) along the $a$~axis are antiferromagnetic and $\nicefrac{1}{5}$ are ferromagnetic.
This model, however, is not supported by the experimental data.
In the zero-field ground state, the situation is reversed: $\nicefrac{4}{5}$ of these couplings are ferromagnetic and only $\nicefrac{1}{5}$ are antiferromagnetic.
The field-induced magnetic structure observed experimentally, characterised by $\mathbf{k}_{\mathrm{1/5}}=(0,0,1/5)$ and lacking a fivefold periodicity along the $a$ axis, preserves ferromagnetic nearest-neighbour couplings along $a$ and therefore requires a much smaller rearrangement of the exchange interactions.
Retaining the fivefold periodicity along the $c$~axis through the \textsc{uduud} spin sequence present in the zero-field ground state yields a structure consistent with the $\nicefrac{1}{5}$ magnetisation plateau while introducing only minimal modifications to the zero-field configuration [Fig.~\ref{fig:MagStructure}(c)].

Please note that only in Fig.~\ref{fig:MagStructure}(a) the sizes of the magnetic moments present the results of zero-field powder neutron diffraction refinements, while for depicting possible configurations in the $\nicefrac{1}{5}$~plateau region, Fig.~\ref{fig:MagStructure}(b) and (c), as well as in the field-polarised state, Fig.~\ref{fig:MagStructure}(d), we presume equal moment sizes for all Nd$^{3+}$ ions.

Based on the diffraction data only in the $(hk0)$ scattering plane, it would have not been possible to establish the field evolution of the in-plane all-in-all-out spin component.
By comparing the high-field magnetisation measurements~\cite{2018_Brunt_RB4} with the moment size obtained from the powder neutron diffraction~\cite{2024_Khalyavin}, one could guess that the in-plane component might survive even in the high-field phase.
A direct access to the peaks of the $(h0l)$ type granted by the horizontal field configuration allowed us to prove this conjecture.

The moment determined from magnetization measurements is about 2~$\mu_{\rm B}$, which is very close to the out-of-plane spin component 1.98(1)~$\mu_{\rm B}$ obtained in the refinement of the neutron diffraction data~\cite{2024_Khalyavin}.
The total moment size (in-plane plus out-of-plane components), coming from the neutron diffraction is 2.55(1)~$\mu_{\rm B}$, impaling that the ``fully'' polarised state (with all the moments pointing along the applied field) is never established even in high applied fields.
Figure~\ref{fig:hk0_field} clearly shows that the peaks (400) and (420) drop in intensity in 33~kOe on a transition to a field-polarised state.
That indicates that the $\mathbf{k}_1=0$ in-plane component changes only at this transition rather than on entering the $\nicefrac{1}{5}$-magnetisation plateau.
Figure~\ref{fig:h0l_field} provides a direct confirmation that the in-plane component survives in applied fields above the 33~kOe transition.
For that reason, labeling the high-field phase as a fully field-polarised phase is incorrect.

Finally, we note the close connection between the in-plane and out-of-plane ($\mathbf{k}_1=0$) spin components, established in Ref.~\cite{2024_Khalyavin} through a trilinear term in the free energy.
This term couples these magnetic order parameters to a non-magnetic order parameter, referred to as the ``hidden'' order parameter. 
Although the latter was not directly observed in the diffraction experiment, its presence was inferred from the critical behaviour of the magnetic order parameters.
The trilinear free-energy term implies that changing the out-of-plane spin component ($\mathbf{k}_1=0$) from antiferromagnetic to ferromagnetic requires a change in the symmetry of either the in-plane spin component or the hidden order parameter.
Since the neutron diffraction data discussed above reveal no change in the in-plane spin component, we conclude that the symmetry of the hidden order parameter changes from GM5$+$ to GM1$-$.
This, in turn, implies a change in the crystal symmetry from monoclinic $P2_1/n$ to tetragonal $P4_212$.

\begin{acknowledgments}
The authors acknowledge financial support from the EPSRC, UK, through Grant EP/M028771/1 and EP/T005963/1.
\end{acknowledgments}

\bibliography{References2}

@article{1972_Buschow,
author = {Buschow, K. H. J.  and Creyghton, J. H. N. },
title = {Magnetic Properties of Rare Earth Tetraborides},
journal = {J. Chem. Phys.},
volume = {57},
number = {9},
pages = {3910-3914},
year = {1972},
doi = {10.1063/1.1678862}}

@article{1973_Creyghton,
  title = {Nuclear Magnetic Resonance of $^{11}\mathrm{B}$ at the Three Boron Sites in Rare-Earth Tetraborides},
  author = {Creyghton, J. H. N. and Locher, P. R. and Buschow, K. H. J.},
  journal = {Phys. Rev. B},
  volume = {7},
  issue = {11},
  pages = {4829--4843},
  numpages = {0},
  year = {1973},
  month = {Jun},
  publisher = {American Physical Society},
  doi = {10.1103/PhysRevB.7.4829},
  url = {https://link.aps.org/doi/10.1103/PhysRevB.7.4829}}

@article{1979_Pfeiffer_ErB4,
title = "The magnetic phase diagram of {ErB}$_4$",
journal = "J. Magn. Magn. Mater.",
volume = "14",
number = "2",
pages = "306 - 308",
year = "1979",
note = "",
issn = "0304-8853",
author = "F. Pfeiffer and W. Schäfer and G. Will and J. Etourneau and R. Georges"}

@article{1981_Shastry,
	author = {Shastry, Sriram B. and Sutherland, Bill},
	Title = {Exact ground state of a quantum mechanical antiferromagnet},
	journal = {Physica B+C"},
	volume = {108},
	number = {1-3},
	pages = {1069 - 1070},
	year = {1981},
	issn = {0378-4363},
	doi = {10.1016/0378-4363(81)90838-X}}

@article{1985_Etourneau_RB4,
author = { J.   Etourneau  and  P.   Hagenmuller },
title = {Structure and physical features of the rare-earth borides},
journal = {Philos. Mag. B},
volume = {52},
number = {3},
pages = {589-610},
year  = {1985},
publisher = {Taylor & Francis}}

@article{2001_Salamakha,
title = "Single crystal investigation of the binary $\mathrm{NdB_4}$ compound",
journal = "J. Alloy. Comp.",
volume = "316",
number = "1",
pages = "L4 - L6",
year = "2001",
issn = "0925-8388",
doi = "https://doi.org/10.1016/S0925-8388(00)01498-5",
url = "http://www.sciencedirect.com/science/article/pii/S0925838800014985",
author = "P. Salamakha and A.P. Concalves and O. Sologub and M. Almeida"}

@article{2006_Blanco_GdB4,
	author = {Blanco, J. A. and Brown, P. J. and Stunault, A. and Katsumata, K. and Iga, F. and Michimura, S.},
	Title = {Magnetic structure of $\mathrm{GdB_4}$ from spherical neutron polarimetry},
  	journal = {Phys. Rev. B},
  	volume = {73},
  	issue = {21},
  	pages = {212411},
  	numpages = {4},
  	year = {2006},
  	month = {Jun},
  	publisher = {American Physical Society},
	doi = {10.1103/PhysRevB.73.212411}}

@article{2008_Okuyama_HoB4,
	author = {Okuyama, Daisuke and Matsumura, Takeshi and Mouri, Takuya and Ishikawa, Naoko and Ohoyama, Kenji and Hiraka, Haruhiro and Nakao, Hironori and Iwasa, Kazuaki and Murakami, Youichi},
	Title = {Competition of Magnetic and Quadrupolar Order Parameters in $\mathrm{HoB_4}$},
	journal = {J. Phys. Soc. Jpn.},
	volume = {77},
	number = {14},
	pages = {044709},
	year = {2008},
	doi = {10.1143/JPSJ.77.044709}}

@article{2008_Siemensmeyer_TmB4,
 	author = {Siemensmeyer, K. and Wulf, E. and Mikeska, H.-J. and Flachbart, K. and Gab\'ani, S. and Mat'a\ifmmode \check{s}\else \v{s}\fi{}, S. and Priputen, P. and Efdokimova, A. and Shitsevalova, N.},
	Title = {Fractional Magnetization Plateaus and Magnetic Order in the {S}hastry-{S}utherland Magnet $\mathrm{TmB_4}$},
  	journal = {Phys. Rev. Lett.},
  	volume = {101},
  	issue = {17},
  	pages = {177201},
 	 numpages = {4},
  	year = {2008},
  	month = {Oct},
  	publisher = {American Physical Society},
  	doi = {10.1103/PhysRevLett.101.177201}}

@article{2009_Moliner,
  title = {Magnetization process in the classical {H}eisenberg model on the {S}hastry-{S}utherland lattice},
  author = {Moliner, M. and Cabra, D. C. and Honecker, A. and Pujol, P. and Stauffer, F.},
  journal = {Phys. Rev. B},
  volume = {79},
  issue = {14},
  pages = {144401},
  numpages = {8},
  year = {2009},
  month = {Apr},
  publisher = {American Physical Society}}

@article{2009_Michimura_TmB4,
	author = { Shinji Michimura and  Akira Shigekawa and  Fumitoshi Iga and  Toshiro Takabatake and  Kenji Ohoyama},
	Title = {Complex Magnetic Structures of a {S}hastry-{S}utherland Lattice $\mathrm{TmB_4}$ Studied by Powder Neutron Diffraction Analysis},
	journal = {J. Phys. Soc. Jpn.},
	volume = {78},
	number = {2},
	pages = {024707},
	year = {2009},
	doi = {10.1143/JPSJ.78.024707}}

@article{2009_Watanuki_NdB4,
author={Watanuki, Ryuta and Kobayashi, Toshihiko and Noguchi, Ryo and Suzuki, Kazuya},
  title={Possible multipolar transition in {NdB}$_4$},
  journal={J. Phys.: Conf. Ser.},
  volume={150},
  number={4},
  pages={042229},
  year={2009}}

@article{2010_Matas,
  author={S Mat'as and K Siemensmeyer and E Wheeler and E Wulf and R Beyer and Th Hermannsdörfer and O Ignatchik and M Uhlarz and K
Flachbart and S Gabáni and P Priputen and A Efdokimova and N Shitsevalova},
  title={Magnetism of rare earth tetraborides},
  journal={J. Phys.: Conf. Ser.},
  volume={200},
  number={3},
  pages={032041},
  year={2010}}

@article{2012_Huang,
  author={W C Huang and L Huo and G Tian and H R Qian and X S Gao and M H Qin and J-M Liu},
  title={Multi-step magnetization of the {I}sing model on a {S}hastry–{S}utherland lattice: a {M}onte {C}arlo simulation},
  journal={J. Phys. - Condens. Mat.},
  volume={24},
  number={38},
  pages={386003},
  year={2012}}

@article{2013_Huo,
   author = "Huo, L. and Huang, W. C. and Yan, Z. B. and Jia, X. T. and Gao, X. S. and Qin, M. H. and Liu, J.-M.",
   title = "The competing spin orders and fractional magnetization plateaus of the classical {H}eisenberg model on {S}hastry-{S}utherland lattice: Consequence of long-range interactions",
   journal = "J. Appl. Phys.",
   year = "2013",
   volume = "113",
   number = "7",
   pages = ""}

@article{2013_Grechnev,
  title = {Exact ground state of the {S}hastry-{S}utherland lattice with classical {H}eisenberg spins},
  author = {Grechnev, Alexei},
  journal = {Phys. Rev. B},
  volume = {87},
  issue = {14},
  pages = {144419},
  numpages = {8},
  year = {2013},
  publisher = {American Physical Society}}

@article{2013_Dublenych,
  title = {Ground states of an Ising model on an extended {S}hastry-{S}utherland lattice and the $1/2$-magnetization plateau in some rare-earth-metal tetraborides},
  author = {Dublenych, Yu. I.},
  journal = {Phys. Rev. E},
  volume = {88},
  issue = {2},
  pages = {022111},
  numpages = {13},
  year = {2013},
  publisher = {American Physical Society}}

@article{2013_Wierschem,
	author = {Wierschem, Keola and Sengupta, Pinaki},
	Title = {Columnar Antiferromagnetic Order and Spin Supersolid Phase on the Extended {S}hastry-{S}utherland Lattice},
	journal = {Phys. Rev. Lett.},
  	volume = {110},
  	issue = {20},
  	pages = {207207},
  	numpages = {5},
  	year = {2013},
  	month = {May},
  	publisher = {American Physical Society},
  	doi = {10.1103/PhysRevLett.110.207207}}

@article{2017_Brunt_HoB4,
	author = {Brunt, D. and Balakrishnan, G. and Wildes, A. R. and Ouladdiaf, B. and Qureshi, N. and Petrenko, O. A.},
	Title = {Field-induced magnetic states in holmium tetraboride},
  	journal = {Phys. Rev. B},
  	volume = {95},
  	issue = {2},
  	pages = {024410},
  	numpages = {8},
  	year = {2017},
  	publisher = {American Physical Society},
  	doi = {10.1103/PhysRevB.95.024410}}

@article{2017_Yamauchi_NdB4,
	author = {H Yamauchi and N Metoki and R. Watanuki and K Suzuki and H Fukazawa and S Chi and J A Fernandez-Baca},
	Title = {Magnetic Structure and Quadrupolar Order Parameter Driven by Geometrical Frustration Effect in $\mathrm{NdB_4}$},
	journal = {J. Phys. Soc. Jpn},
	volume = {86},
	number = {4},
	pages = {044705},
	year = {2017},
	doi = {10.7566/JPSJ.86.044705}}

@phdthesis{Brunt_thesis,
	Author = {Daniel Brunt},
	School = {The University of Warwick},
	Title = {Magnetic Properties of Frustrated {S}hastry-{S}utherland Magnets; Rare Earth Tetraborides},
	Year = 2017}

@article{2018_Metoki_NdB4,
 	 author = {Metoki, N. and Yamauchi, H. and Matsuda, M. and Fernandez-Baca, J. A. and Watanuki, R. and Hagihala, M.},
	Title = {Polarized neutron scattering study of the multiple order parameter system $\mathrm{NdB_4}$},
   	journal = {Phys. Rev. B},
  	volume = {97},
  	issue = {17},
  	pages = {174416},
  	numpages = {10},
  	year = {2018},
  	month = {May},
  	publisher = {American Physical Society},
  	doi = {10.1103/PhysRevB.97.174416}}

@article{2018_Brunt_RB4,
    author = {Brunt, Daniel and Balakrishnan, Geetha and Mayoh, Daniel and Lees, Martin R. and Gorbunov, Denis I. and Qureshi, Navid and Petrenko, Oleg A.},
	title = {Magnetisation process in the rare earth tetraborides, $\mathrm{NdB_4}$ and $\mathrm{HoB_4}$},
	year = {2018},
	month = {12},
	pages = {232},
	volume = {8},
	journal = {Sci. Rep.},
  	doi = {10.1038/s41598-017-18301-1}}

@Article{2019_Brunt_Crystal,
	Author = {Brunt, Daniel and Hatnean, Monica Ciomaga and Petrenko, Oleg A. and Lees, Martin R. and Balakrishnan, Geetha},
  	Title = {Single-Crystal Growth of Metallic Rare-Earth Tetraborides by the Floating-Zone Technique},
	Journal = {Crystals},
	Year = {2019},
	Volume = {9},
	Number = {4},
	pages = {211},
  	doi = {10.3390/cryst9040211}}

@article{2019_Lancon,
  	author = {Lan\ifmmode \mbox{\c{c}}\else \c{c}\fi{}on, D. and Scagnoli, V. and Staub, U. and Petrenko, O. A. and Ciomaga Hatnean, M. and Canevet, E. and Sibille, R. and Francoual, S. and Mardegan, J. R. L. and Beauvois, K. and Balakrishnan, G. and Heyderman, L. J. and R\"uegg, Ch. and Fennell, T.},
	Title = {Evolution of field-induced metastable phases in the {S}hastry-{S}utherland lattice magnet $\mathrm{TmB_4}$},
  	journal = {Phys. Rev. B},
  	volume = {102},
  	issue = {6},
  	pages = {060407},
  	numpages = {6},
  	year = {2020},
  	month = {Aug},
  	publisher = {American Physical Society},
  	doi = {10.1103/PhysRevB.102.060407}}

@article{2021_Ohlendorf,
  title = {Magnetoelastic coupling and Gr\"uneisen scaling in ${\mathrm{NdB}}_{4}$},
  author = {Ohlendorf, R. and Spachmann, S. and Fischer, L. and Dey, K. and Brunt, D. and Balakrishnan, G. and Petrenko, O. A. and Klingeler, R.},
  journal = {Phys. Rev. B},
  volume = {103},
  issue = {10},
  pages = {104424},
  numpages = {6},
  year = {2021},
  month = {Mar},
  publisher = {American Physical Society},
  doi = {10.1103/PhysRevB.103.104424},
  url = {https://link.aps.org/doi/10.1103/PhysRevB.103.104424}}

@article{2022_Qureshi,
  title = {Possible stripe phases in the multiple magnetization plateaus in {TbB}$_4$ from single-crystal neutron diffraction under pulsed high magnetic fields},
  author = {Qureshi, N. and Bourdarot, F. and Ressouche, E. and Knafo, W. and Iga, F. and Michimura, S. and Regnault, L.-P. and Duc, F.},
  journal = {Phys. Rev. B},
  volume = {106},
  issue = {9},
  pages = {094427},
  numpages = {8},
  year = {2022},
  month = {Sep},
  publisher = {American Physical Society},
  doi = {10.1103/PhysRevB.106.094427},
  url = {https://link.aps.org/doi/10.1103/PhysRevB.106.094427}}

@article{2023_Ohlendorf,
  title = {Magnetic phase diagram and magnetoelastic coupling {NdB}$_4$ studied by high-resolution capacitance dilatometry up to 35~{T}},
  author = {Ohlendorf, R. and Spachmann, S. and Fischer, L. and Carstens, F. L. and Brunt, D. and Balakrishnan, G. and Petrenko, O. A. and Klingeler, R.},
  journal = {Phys. Rev. B},
  volume = {108},
  issue = {22},
  pages = {224411},
  numpages = {10},
  year = {2023},
  month = {Dec},
  publisher = {American Physical Society},
  doi = {10.1103/PhysRevB.108.224411},
  url = {https://link.aps.org/doi/10.1103/PhysRevB.108.224411}}

@article{2024_Khalyavin,
  title = {Magnetic ground state of {NdB}$_4$: {I}nterplay between anisotropic exchange interactions and hidden order on a {S}hastry-{S}utherland lattice},
  author = {Khalyavin, D. D. and Brunt, D. and Qureshi, N. and Wildes, A. R. and Ouladdiaf, B. and Johnson, R. D. and Balakrishnan, G. and Petrenko, O. A.},
  journal = {Phys. Rev. B},
  volume = {109},
  issue = {22},
  pages = {L220411},
  numpages = {7},
  year = {2024},
  month = {Jun},
  publisher = {American Physical Society},
  doi = {10.1103/PhysRevB.109.L220411},
  url = {https://link.aps.org/doi/10.1103/PhysRevB.109.L220411}}

@article{1988_Wilkinson,
author = "Wilkinson, C. and Khamis, H. W. and Stansfield, R. F. D. and McIntyre, G. J.",
title = "{Integration of single-crystal reflections using area multidetectors}",
journal = "J. Appl. Crystallogr.",
year = "1988",
volume = "21",
number = "5",
pages = "471--478"}

@article{mag2pol,
	Author = {N. Qureshi},
	Journal = {J. Appl. Cryst.},
	Pages = {175},
	Title = {Mag2Pol: A program for the analysis of spherical neutron polarimetry, flipping ratio and integrated intensity data},
	Volume = {52},
	Year = {2019}}

@article{2004_Hardy,
	Author = {Hardy, V. and Lees, M. R. and Petrenko, O. A. and Paul, D. McK. and Flahaut, D. and H\'ebert, S. and Maignan, A.},
	Doi = {10.1103/PhysRevB.70.064424},
	Issue = {6},
	Journal = {Phys. Rev. B},
	Month = {Aug},
	Numpages = {7},
	Pages = {064424},
	Publisher = {American Physical Society},
	Title = {Temperature and time dependence of the field-driven magnetization steps in {C}a$_3${C}o$_2${O}$_6$ single crystals},
	Url = {http://link.aps.org/doi/10.1103/PhysRevB.70.064424},
	Volume = {70},
	Year = {2004}}

@misc{D10_2026,
	note_ = {Raw files of the neutron diffraction data collected on D10 are available at https://doi:10.5291/ILL-DATA.5-41-1351},
	author = {Oleg Petrenko and others},
	title = {Field-induced magnetic order in {N}d{B}$_4$},
	year = {2026},
	url = {https://doi:10.5291/ILL-DATA.5-41-1351},
	note = {https://doi:10.5291/ILL-DATA.5-41-1351}}

@misc{D10_2016, 
	note_ = {Raw files of the neutron diffraction data collected on D10 are available at https://doi:10.5291/ILL-DATA.5-41-871},
	author = {Oleg Petrenko and others},
	title = {Investigation into new incommensurate reflections found in the frustrated magnet, {N}d{B}$_4$},
	year = {2016},
	url = {https://doi:10.5291/ILL-DATA.5-41-871},
	note = {https://doi:10.5291/ILL-DATA.5-41-871}}

@misc{D10_2014, 
	note_ = {Raw files of the neutron diffraction data collected on D10 are available at https://doi:10.5291/ILL-DATA.5-41-762},
	author = {Oleg Petrenko and others},
	title = {Determination of the magnetic structure of the {S}hastry-{S}utherland Magnet Neodymium Tetraboride},
	year = {2014},
	url = {https://doi:10.5291/ILL-DATA.5-41-762},
	note = {https://doi:10.5291/ILL-DATA.5-41-762}}
\end{document}